\documentstyle[preprint,prabib,aps]{revtex}
\begin{document}
\title{Multidimensional Gravity on the Principal
Bundles}
\author{Dzhunushaliev V.D.
\thanks{E-mail address: dzhun@freenet.bishkek.su}}
\address {Theoretical Physics Department \\
          Kyrgyz State National University, Bishkek, 720024, 
Kyrgyzstan\\
Home address: mcr.Asanbai, d.25, kw.24, Bishkek, 
720060, Kyrgyzstan.\\
Telephone: (3312) 46 57 45
}
\maketitle
\newpage
\begin{abstract}
The multidimensional gravity on the total space 
of principal bundle is considered. In this theory 
the gauge fields arise as nondiagonal components of 
multidimensional metric. The spherically symmetric 
and cosmology solutions for gravity on SU(2) principal 
bundle are obtained. The static spherically symmetric 
solution is wormhole-like solution located between 
two null surfaces, in contrast to 4D Einstein-Yang-Mills 
theory where corresponding solution (black hole) located 
outside of event horizon. Cosmology solution 
(at least locally) has the bouncing off effect for 
spatial dimensions. In spirit of Einstein  
these solutions are vacuum solutions without matter. 
\end{abstract}

\section{Introduction}
As well known the Yang-Mills gauge field is the geometrical object: 
a connection on the principal bundle. The base 
of this bundle is spacetime 
and the fibres are structural group. If we include gravity 
then it acts only on the base of this bundle and this is the 
Einstein-Yang-Mills theory. The simplest extension of such 4D 
gravity is assumption that the gravity acts on total 
space of the principal bundle. The relation between 
two these theories give the following theorem 
\cite{per}-\cite{sal}: 
\par
Let $G$ group  be  the
fibre of principal  bundle.  Then  there  is  the  one-to-one
correspondence between $G$-invariant metrics on the  total  
space ${\cal X}$
and the triples $(g_{\mu \nu }, A^{a}_{\mu }, h\gamma _{ab})$. 
Where $g_{\mu \nu }$ is Einstein's pseudo  -
Riemannian metric on the base; $A^{a}_{\mu }$ is gauge field 
of the $G$ group ( the nondiagonal components of multidimensional 
metric); $h\gamma _{ab}$  is 
symmetric metric on the fibre. 
\par 
The such multidimensional (MD) gravity differs from standard 
MD gravity by following manner:
\begin{enumerate}
\item
The extremal dimensions (fibres of bundle) are not equivalent 
to spacetime dimensions, as they make up group.
\item
Any physical fields on the total space of bundle 
(including MD metric) can depend only on base (spacetime) 
coordinates. 
\item 
In this vacuum theory the gauge fields appears by natural way as a 
nondiagonal components of MD metric. 
\end{enumerate}
In standard MD gravity the gauge field is added as an 
external matter field. See, for example \cite{br1}-\cite{br2}, 
where the spherically symmetric and cosmological 
solutions are obtained in MD gravity coupling with 
generalized Maxwell field. In \cite{rei} an inhomogeneous 
multidimensional 
cosmological model with a higher dimensional  space-time manifold 
$M=M_0\times\prod_{i=1}^n M_i$ ($n\geq 1$) are investigated
under dimensional reduction to tensor-multi-scalar theories. 
($M_0$ is Einstein's spacetime, $M_i$ are internal spaces). 

\section{Gravity equations}

We note that the metric on the fibre can have only following view: 
\begin{equation}
ds^{2}_{fiber} = h(x^{\mu }) \sigma ^{a}\sigma _{a},
\label{2-1}
\end{equation}
where conformal factor $h(x^{\mu })$ depends only on spacetime 
coordinates $x^{\mu }; \mu =0,1,2,3; \sigma _{a}=\gamma _{ab}\sigma ^{b}; 
\gamma _{ab}$  is  Euclidean  metric; $a=4,5,\ldots N$ index on 
fibre (internal space). This follows from the fact that 
the fibre is a symmetrical space (gauge group). 
$\sigma ^{a}$  are  one-form satisfies Maurer - Cartan structure 
equations: 
\begin{equation}
d\sigma ^{a} = f^{a}_{bc}\sigma ^{b}\wedge\sigma ^{c},
\label{2-3}
\end{equation}
where $f^{a}_{bc}$ is a structural constant of gauge group. 
Thus, MD metric on  the  total  space
can be written in the following view: 
\begin{equation}
ds^{2} = ds^{2}_{fibre} + 2 G_{A\mu } dx^{A} dx^{\mu },
\label{2-4}
\end{equation}
where $A=0,1,\ldots ,N$  is  multidimensional  index 
on the total space. 
\par
Hence we have only following independent degrees of freedom: 
conformal factor $h(x^\mu)$ and MD metric $G_{A\mu}$. 
Varying with respect to these variables leads to the following 
gravity equations: 
\begin{eqnarray}
R_{A\mu } -{1\over 2}G_{A\mu }R = 0,
\label{2-2-1}\\
R^{a}_{a} = R^{4}_{4} + R^{5}_{5} + R^{6}_{6} = 0.
\label{2-2-2}
\end{eqnarray}
These equations are Einstein's MD equations for gravity on the 
principal bundle in vacuum. 
Below we consider two cases: spherically symmetric 
and cosmology solutions. 

\section{Wormhole-like solutions}
\subsection{U(1) case}

We remind the solution for $5D$ Kaluza - Klein's theory derived
in \cite{dzh1}. The metric is:
\begin{equation}
ds^{2} = e^{2\nu (r)}dt^{2} - e^{2\psi (r)}(d\chi  - 
\omega (r)dt)^2 - dr^{2}-
a^2(r)(d\theta ^{2} + \sin ^{2}\theta  d\varphi ^2),
\label{3-1}
\end{equation}
 where $\chi $ is the 5th supplementary coordinate; 
$r,\theta ,\varphi $ are $3D$  polar coordinates; 
$t$ is the time. The solution of $5D$ Einstein's equations is:
\begin{eqnarray}
a^{2} & = & r^{2}_{0} + r^{2},
\label{3-2-1}\\
e^{-2\psi } = e^{2\nu } & = & {2r_{0}\over q}{r^{2}_{0}+r^{2}
\over r^{2}_{0}-r^{2}},
\label{3-2-2}\\
\omega & = & 4r^{2}_{0}\over q}
{r\over {r^{2}_{0} - r^{2}}.
\label{3-2-3}
\end{eqnarray}
This solution is the wormhole-like object located between 
two null surfaces $(r = \pm r_0)$. We note that this solution 
is nonsingular in $|r| \le r_0$. Really, determinant of 
this MD metrics is equal to: 
\begin{equation}
\det (G_{AB}) =  \sin ^2\theta (r_0^2 + r^2)^2
\label{3-3}
\end{equation}
this is indirect confirmation that multidimensional 
metrics doesn't have singularity on null surface $r=\pm r_0$. 
Also we can say that from Einstein's equations follows 
that $R_B^A = 0$ and hence $R_A^B R^A_B < \infty$ ($A,B=0,1,2,3,4$). 
At last it can shown that invariant $R^A_{BCD}R_A^{BCD}< \infty$.
\par 
Such solution we can name as cutting off wormhole (WH) in 
contrast to standard WH joining two asymptotical flat 
regions. 

\subsection{SU(2) monopole-like case}

We can introduce the Euler's angles 
$\alpha ,\beta , \gamma$ on fibre ($SU(2)$ group).  Then
one-forms $\sigma ^{a}$ can be written as a follows:
\begin{eqnarray}
\sigma ^{1} & = & {1\over 2}
(\sin \alpha d\beta - \sin \beta \cos \alpha d\gamma ),
\label{3-4-1}\\
\sigma ^{2} & = & -{1\over 2}(\cos \alpha d\beta +
\sin \beta \sin \alpha d\gamma ),
\label{3-4-2}\\
\sigma ^{3} & = & {1\over 2}(d\alpha +\cos \beta d\gamma ),
\label{3-4-3}
\end{eqnarray}
where $0\le \beta \le \pi , 0\le \gamma \le 2\pi , 0\le \alpha \le 4\pi $. 
We see a solution of the form:
\begin{equation}
ds^{2} = e^{2\nu (r)}dt^{2}  - r^{2}_{0}e^{2\psi (r)}\sum^{3}_{a=1}
\left (\sigma ^{a} - A^{a}_{\mu }(r)dx^{\mu }\right )^{2} -
dr^{2} - a^{2}(r)\left (d\theta ^{2} + 
\sin ^{2}\theta d\varphi ^2\right ) .
\label{3-6}
\end{equation}
We choose the ``potentials'' $A^{a}_{\mu }$ in following 
monopole-like form:
\begin{eqnarray}
A^{a}_{\theta } & = & {1\over 2}(f(r)+1)\{ \sin \varphi ;-\cos \varphi ;
0\} ,
\label{3-7-1}\\
A^{a}_{\varphi } & = & {1\over 2}(f(r)+1)
\sin \theta \{\cos \varphi \cos \theta ;
\sin \varphi \cos \theta ;-\sin \theta \},
\label{3-7-2}\\
A^{a}_{t} & = & v(r)\{ \sin \theta \cos \varphi ;
\sin \theta \sin \varphi ;\cos \theta \},
\label{3-7-3}
\end{eqnarray}
Let  us  introduce  tetrads $e^{\bar{A}}_{A}$:
\begin{eqnarray}
ds^{2} & = & \eta _{\bar{A}\bar{B}}\Sigma ^{\bar{A}}
\Sigma ^{\bar{B}},
\label{3-8-1}\\
\Sigma ^{\bar{A}} & = & e^{\bar{A}}_{A} dx^{A},
\label{3-8-2}
\end{eqnarray}
where $\bar{A},\bar{B}=0,1,\ldots ,6$ are  tetrads  indexes; 
$\eta _{\bar{A}\bar{B}}$  is $7D$  Minkowski
metric. The input equations are written  below  in  the  following
form:
\begin{eqnarray}
R_{\bar{A}\mu } & = & 0,
\label{3-9-1}\\
R^{a}_{a} & = & 0.
\label{3-9-2}
\end{eqnarray}
$7D$ gravity equations become:
\begin{eqnarray}
\nu ''  + {\nu '}^2 + 3\nu ' \psi '  
+ 2{a' \nu ' \over a} -
{r^{2}_{0} \over 2} e^{2(\psi -\nu )}{v'}^2 -
{r^{2}_{0}\over a^{2}} v^{2}f^{2}e^{2(\psi -\nu )}  =  0,
\label{3-10-1}\\
\nu ''  + {\nu '}^2 + 3\psi ''  + 
3{\psi '}^2 + 2{a'' \over a} -
{r^{2}_{0}\over 2} e^{2(\psi -\nu )}{v'}^2 +
{r^{2}_{0}\over 4a^2} {f'}^2e^{2\psi }  =  0,
\label{3-10-2}\\
{a'' \over a} + {a'\over a}(\nu ' + 3\psi ' ) + 
{{a'}^2\over a^2}  - {1\over a^2}  + 
{r^{2}_{0}\over 8a^2} e^{2\psi }{f'}^2 -
{r^{2}_{0}\over 2a^2} v^2 f^2 e^{2(\psi -\nu )} +
{r^{2}_{0}\over 8a^{4}}\left (f^2 -1\right )^2  =  0,
\label{3-10-3}\\
\psi ''  + 3{\psi '}^2 + 
2{a' \psi '\over a} + \psi ' \nu '  +
{r^{2}_{0}\over 6} e^{2(\psi -\nu )} {v'}^2 -
{2\over r^{2}_{0}} e^{-2\psi } - 
{r^{2}_{0}{f'}^2\over 12a^2} e^{2\psi } + 
\nonumber\\
{r^{2}_{0}\over 3}{v^2 f^2\over a^2}e^{2(\psi -\nu )} -
{r^{2}_{0}\over 24a^4} \left (f^2 -1\right )^2  =  0,
\label{3-10-4}\\
f''  + f' (\nu ' + 5\psi ') + 
2v^2 f e^{-2\nu }  = 
{f\over 2a^{2}}(f^2 -1),
\label{3-10-5}\\
v''  - v'(\nu ' - 5\psi ' - 2{a'\over a})  =  2{v\over a}f^{2},
\label{3-10-6}
\end{eqnarray}
here the Eq's(\ref{3-10-5}) and (\ref{3-10-6}) are ``Yang  -  Mills''  
equations  for nondiagonal  components  of  the MD 
metric. For simplicity we consider $f=0$ case. 
This means that  we  have ``color electrical'' 
field $A^{a}_{i}$ only (i=1,2,3). 
In this case it  is  easy  to integrate 
Eq.(\ref{3-10-6}): 
\begin{equation}
v'  ={q\over r_{0}a^{2}} e^{\nu -5\psi },
\label{3-11}
\end{equation}
where $q$ is the constant of  the  integration  (``color  electrical''
charge). Let us examine the most interesting case when the  linear
dimensions  of  fibres $r_{0}$  are  vastly  smaller  than  the  space
dimension $a_{0}$ and ``charge'' $q$ is sufficiently large:
\begin{equation}
\left ({q\over a_{0}}\right )^{1/2} \gg 
\left ({a_0\over r_{0}}\right )^2 \gg  1,
\label{3-12}
\end{equation}
 where $a_{0}=a(r=0)$ is the throat of the WH.
\par
On this approximation we deduce the equations system:
\begin{eqnarray}
\nu ''  + {\nu '}^2 + 
3\nu ' \psi '  + 2{a' \nu '\over a}  - {q^2\over 2a^4} e^{-8\psi } 
 &  = & 0,
\label{3-13-1}\\
\nu ''  + {\nu '}^2 + 3\psi ''  + 3{\psi '}^2 + 
2{a'' \over a} - {q^2\over 2a^4} e^{-8\psi }
&   = & 0,
\label{3-13-2}\\
\psi ''  + 3{\psi '}^2 + 2{a' \psi '\over a} + 
\psi ' \nu '  + {q^2\over 6a^4} e^{-8\psi } 
&  = & 0,
\label{3-13-3}\\
{a'' \over a} + {a'\over a}(\nu ' + 3\psi ' ) + 
{{a'}^2\over a^2}  - {1\over a^2} 
&  = & 0.
\label{3-13-4}
\end{eqnarray}
This system has the following solution:
\begin{eqnarray}
\nu & = & -3\psi,
\label{3-14-1}\\
a^2 & = & a^2_0 + r^2,
\label{3-14-2}\\
e^{-{4\over 3}\nu } & = & {q\over 2a_0}
\cos \left (\sqrt {8\over 3}\arctan {r\over a_0} 
     \right ),
\label{3-14-3}\\
v & = & \sqrt 6{a_{0}\over r_0 q}
\tan \left (\sqrt {8\over 3} \arctan {r\over a_0}
     \right ).
\label{3-14-4}
\end{eqnarray}
Let us define value $r$ in which  metric  has  null surfaces.  From
condition:
\begin{equation}
G_{tt}(r_g) = e^{2\nu (r_g)} - 
r^2_0 e^{2\psi (r)}\sum^3_{a=1}\left (A^a_t(r_g) \right )^{2} = 0
\label{3-15}
\end{equation}
it follows that:
\begin{equation}
{r_g\over a_0} = 
\tan \left (\sqrt {3\over 8}\arcsin \sqrt {2\over 3} 
     \right ) \approx  0.662.
\label{3-16}
\end{equation}
 It   is   easy   to   verify   that $\exp (2\nu )=0$ 
$(\exp (2\psi =\infty )$    by
$r/a_{0}=\tan (\pi \sqrt{3/32})\approx 1.434$. 
This  value  lies  beyond  the  null 
surfaces. This means that the small terms in (\ref{3-10-1}), 
(\ref{3-10-4}) will stay also small even near the null surfaces.

\section{SU(2) cosmology solution}

Analogous to spherically symmetric metric (\ref{3-6}) we will 
search the MD cosmology metric on the total space as follows: 
\begin{eqnarray} 
ds^2 & = & dt^2 - b^2(t)\sum^{3}_{a=1}
\left (\sigma ^{a} - A^{a}_{\mu }(r)dx^{\mu }\right )^{2} -
a^{2}(t)\left (\sigma _x^2 + \sigma _y^2 + 
\sigma _z^2 \right ),
\label{4-1-1}\\
\sigma _x & = & \frac{1}{2}(\sin\psi d\theta - 
\sin\theta \cos\psi d\varphi ),
\label{4-1-2}\\
\sigma _y &= & \frac{1}{2}(-\cos\psi d\theta - 
\sin\theta \sin\psi d\varphi ),
\label{4-1-3}\\
\sigma _z &= & \frac{1}{2}(d\psi + \cos\theta d\varphi ).
\label{4-1-4}
\end{eqnarray}
here $\psi , \theta ,\varphi$ are Euler's angles on 
$S^3$ (3D sphere is spacelike section of Universe). 
We write down the nondiagonal components of MD metric 
in following instanton-like form: 
\begin{eqnarray}
A^a_\chi & = & \frac{1}{4}\left \{ -\sin\theta \cos\varphi ;
-\sin\theta \sin\varphi ;\cos\theta \right \}
(f(t) - 1),
\label{4-2-1}\\
A^a_\theta & = & \frac{1}{4}\left \{ -\sin\varphi ;
-\cos\varphi ;0\right \}(f(t) - 1),
\label{4-2-2}\\
A^a_\varphi & = & \frac{1}{4}\left \{0;0;1\right \}
(f(t) - 1).
\end{eqnarray}
After substitution to initial gravity equations 
(\ref{3-9-1})-(\ref{3-9-2}) we have: 
\begin{eqnarray}
\frac{\ddot a}{a} = \frac{2}{a^2} + \frac{4}{b^2} + 
2\frac{\dot a ^2}{a^2} + 4\frac{\dot b^2}{b^2} + 
9\frac{\dot a\dot b}{ab} & - & \frac{3}{8}\dot f^2 \frac{b^2}{a^2}, 
\label{4-3-1}\\
\frac{\ddot b}{b} = -\frac{2}{a^2} - \frac{4}{b^2} - 
2\frac{\dot a ^2}{a^2} - 4\frac{\dot b^2}{b^2} - 
9\frac{\dot a\dot b}{ab} & + & \frac{1}{4}\dot f^2 \frac{b^2}{a^2}, 
\label{4-3-2}\\
\ddot f = -\dot f \left (5\frac{\dot b}{b} + 
\frac{\dot a}{a}\right ) & + & \frac{2}{a^2}f\left (1 - f^2\right ),
\label{4-3-3}\\
\frac{1}{a^2} + \frac{1}{b^2} + \frac{\dot a^2}{a^2} + 
\frac{\dot b^2}{b^2} + 3\frac{\dot a\dot b}{ab} - 
\frac{1}{16}\dot f^2\frac{b^2}{a^2} - 
\frac{1}{16}\frac{b^2}{a^4}\left (f^2 - 1\right )^2 & = & 0.
\label{4-3-4}
\end{eqnarray}
These equations have the following interesting properties. 
In 4D case Friedman-Walker-Robertson (FRW) 
solutions is not the solutions with bouncing off (they don't 
have even though local minimum). Let us consider the 
time moment in which the all functions 
$a(t), b(t), f(t)$ have the local extreme and analyze 
this extremum. In this point we have the following 
expression for $\ddot a_0, \ddot b_0$ and $\ddot f_0$: 
\begin{eqnarray}
\frac{\ddot a_0}{a_0} & =  & \frac{2}{a^2_0} + \frac{4}{b^2_0},
\label{4-4-1}\\
\frac{\ddot b_0}{b_0} & =  & -\frac{2}{a^2_0} - \frac{4}{b^2_0},
\label{4-4-2}\\
\ddot f_0 & = & \frac{2}{a^2_0}f_0\left (1 - f^2_0\right ),
\end{eqnarray}
here sign 0 indicate that the value of corresponding 
function at $t=0$ is given. 
From these equations we see that 3D Universe (time section 
of 4D base of principal bundle) has, at least locally, 
a bouncing off effect in contrast with 
4D case. This 
leads from the fact that the effective 4D stress-energy tensor 
derived from metric on extremal dimension evidently violate 
the strong energy conditions. The total space of principal 
bundle behaves as MD Kasner's Universe with expanding 
space dimensions and contracting extremal dimensions 
(at least locally). But unlike to 
standard MD gravity the space coordinates can be only expand and 
respectively the extremal dimensions (fibre of principal bundle) 
only shrink.
\par
The general solution of Eq's (\ref{4-3-1}-\ref{4-3-4}) 
has a singularity. Let us investigate the behaviour of functions 
$a(t), b(t), f(t)$ near this singularity. We will search 
solution in this region in the following form: 
\begin{eqnarray}
a(t) & \approx & a_\infty(t-t_0)^\alpha,
\label{4-5-1}\\
b(t) & \approx & b_\infty(t-t_0)^\beta,
\label{4-5-2}\\
f(t) & \approx & f_\infty + f_1(t-t_0)^\gamma,
\label{4-5-3}
\end{eqnarray}
$a_\infty , b_\infty , f_\infty ,f_1$ are some constants. 
The simple calculations give us the following results: 
\begin{eqnarray}
\alpha & = & \frac{1 \mp \sqrt 5}{6},
\label{4-6-1}\\
\beta & = & \frac{1 \pm \sqrt 5}{6},
\label{4-6-2}\\
\gamma & = & \frac{5 \pm \sqrt 5}{3},
\label{4-6-3}
\end{eqnarray}
The initial equations are very difficult for analytical 
investigations and hence we solve these equations only 
numerically with following bouncing off initial conditions: 
\begin{eqnarray}
a(0) = a_0,\quad \dot a(0) = 0,
\nonumber\\
b(0) = b_0,\quad \dot b(0) = 0,
\nonumber\\
f(0) = f_0,\quad \dot f(0) = 0,
\end{eqnarray}
Without loss of generality we can take $a_0=1$. 
The condition for $b_0$ follows from initial conditions 
Eq. (\ref{4-3-4}): 
\begin{equation}
\left (\frac{b_0}{a_0}\right )^2 = 8\frac{1 + \sqrt{1 + 
\frac{\left (1 - f_0^2\right )^2}{4}}}{\left (1 - f_0^2\right )^2}
\end{equation}
Thus, this system has only own independent 
parameter $f_0$. The typically solution of these equations 
is presented on Fig.1-3. From these Fig's we see 
that our solution has a local bouncing off effect by 
$t=0$ and singularity by some $t_0$. 

\section{Conclusion}

Finally, we can to say following: 4D Einstein-Yang-Mills 
theory and corresponding MD gravity on the principal 
bundle conform to each another in some sense. But the 
dynamic of these theories is sufficiently another. 
In static spherically symmetric case the 4D Einstein-Yang-Mills 
theory has solution outside of event horizon 
(black hole filled by Yang-Mills gauge field 
Ref's (\cite{bar}) whereas analogously solution in MD 
gravity there is under null surfaces. Further, 4D FRW 
solution doesn't have the bouncing off but MD gravity 
on the principal bundle has (at least locally) 
bouncing off effect in general solution. Most likely 
this take place from the fact that the MD gravity on 
principal bundle can violate the energy conditon. 

\section{Acknowledzments}

I am very gratefull to DAAD for stipendium and invitation 
to Freie Universitaet Berlin.

\end{document}